\documentclass[a4paper,UKenglish,cleveref, autoref, thm-restate]{lipics-v2021}
\usepackage{graphicx,amsmath,amssymb,amsthm,mathtools,url}
\usepackage{hyperref,lineno,verbatim,soul,tabulary}
\newcolumntype{C}[1]{>{\centering\arraybackslash}p{#1}}
\usepackage{xcolor}
\usepackage{soul}
\newcommand{\mc}[2]{\colorbox{#1}{$\displaystyle #2$}}

\theoremstyle{definition}
\newtheorem{thm}{Theorem}[section]
\newtheorem{dfn}[thm]{Definition}
\newtheorem{pro}[thm]{Problem}
\newtheorem{prop}[thm]{Proposition}

\newtheorem{lem}[thm]{Lemma}
\newtheorem{exa}[thm]{Example}

\newcommand{\Z}{\mathbb Z}
\newcommand{\R}{\mathbb R}
\newcommand{\al}{\alpha}
\newcommand{\be}{\beta}

\newcommand{\de}{\delta}

\newcommand{\La}{\Lambda}
\newcommand{\si}{\sigma}
\newcommand{\ep}{\varepsilon}
\newcommand{\ph}{\varphi}

\newcommand{\TM}{\mathrm{TM}}
\newcommand{\SO}{\mathrm{SO}}
\newcommand{\Or}{\mathrm{O}}

\newcommand{\DM}{\mathrm{DM}}

\newcommand{\CDM}{\mathrm{CDM}}
\newcommand{\MCM}{\mathrm{MCM}}

\newcommand{\TVI}{\mathrm{TVI}}
\newcommand{\VNM}{\mathrm{VNM}}
\newcommand{\TNM}{\mathrm{TNM}}
\newcommand{\RNM}{\mathrm{RNM}}
\newcommand{\INM}{\mathrm{INM}}
\newcommand{\IT}{\mathrm{IT}}
\newcommand{\DL}{\mathrm{DL}}

\newcommand{\SDL}{\mathrm{SDL}}
\newcommand{\NDL}{\mathrm{NDL}}
\newcommand{\Elm}{\mathrm{Elm}}
\newcommand{\lcm}{\mathrm{lcm}}

\newcommand{\sym}{\mathrm{Sym}}

\newcommand{\bs}{\hfill $\blacksquare$}

\newcommand{\vl}{\,:\,}

\title{Exactly computable and continuous metrics on isometry classes of finite and 1-periodic sequences} %TODO Please add
\titlerunning{Computable metrics on isometry classes of finite and 1-periodic sequences} %TODO optional, please use if title is longer than one line

\author{Vitaliy Kurlin}{Department of Computer Science, University of Liverpool, Liverpool, United Kingdom}{vitaliy.kurlin@liverpool.ac.uk}{https://orcid.org/0000-0001-5328-5351}{EPSRC grant `Application-driven Topological Data Analysis', EP/R018472/1}

\authorrunning{Vitaliy Kurlin} %TODO mandatory. First: Use abbreviated first/middle names. Second (only in severe cases): Use first author plus 'et al.'

%TODO mandatory: Please choose ACM 2012 classifications from https://dl.acm.org/ccs/ccs_flat.cfm 

%TODO mandatory; please add comma-separated list of keywords

\nolinenumbers %uncomment to disable line numbering

\hideLIPIcs  %uncomment to remove references to LIPIcs series (logo, DOI, ...), e.g. when preparing a pre-final version to be uploaded to arXiv or another public repository

\EventEditors{John Q. Open and Joan R. Access}
\EventNoEds{2}
\EventLongTitle{42nd Conference on Very Important Topics (CVIT 2016)}
\EventShortTitle{CVIT 2016}
\EventAcronym{CVIT}
\EventYear{2016}
\EventDate{December 24--27, 2016}
\EventLocation{Little Whinging, United Kingdom}
\EventLogo{}
\SeriesVolume{42}
\ArticleNo{23}
\begin{document}

\maketitle

\begin{abstract}
The inevitable noise in real measurements motivates the problem to continuously quantify the similarity between rigid objects such as periodic time series and proteins given by ordered points and considered up to isometry maintaining inter-point distances.
The past work produced many Hausdorff-like distances that have slow or approximate algorithms due to minimizations over infinitely many isometries.
For finite and 1-periodic sequences under isometry in any high-dimensional Euclidean space, we introduce continuous metrics with faster algorithms.
The key novelty in the periodic case is the continuity of new metrics under perturbations that change the minimum period.
\end{abstract}

%ToDo ex 5.2+5.4 on 4-point sequences (figures), proteins in section 1, no URLs in biblio, example on swapping points with close time projections, references to bottleneck drawbacks

\newpage

%1========================
\section{Motivations, problem statement, and overview of new results}
\label{sec:intro}

We start from periodic sequences in $\R$, which will be later extended to higher dimensions.

\begin{dfn}[periodic sequences in $\R$]
\label{dfn:periodic_sequence}
A \emph{periodic sequence} $S=\{p_1,\dots,p_m\}+l\Z$ is defined by a finite \emph{motif} of points $p_1,\dots,p_m$ in a period interval $[0,l)$ of a length $l>0$ as the infinite sequence $p(i+mj)=p_i+jl$ indexed by $i+mj$, where $i=1,\dots,m$ and $j\in\Z$.
\bs
\end{dfn}

Any periodic sequence $S\subset\R$ is infinite in both directions, though all results below can be adapted to 1-directional sequences.
A period interval $[0,l)$ excludes the endpoint $l$, which is equivalent to $0$ by a shift by the period $l$, so any point $p_i\in[0,l)$ is counted once.
\medskip

The set of half-integers is the periodic sequence $\{0,\frac{1}{2}\}+\Z$, which can be also defined as $\{0\}+\frac{1}{2}\Z$.
A period interval $[0,l)$ is \emph{minimal} for a given sequence $S=\{p_1,\dots,p_m\}+l\Z$ if $S$ can not be represented by a shorter period interval.
The points $p_1,\dots,p_m$ are naturally ordered in $[0,l)$.
Fixing a minimum period interval $[0,l)$ at the origin in the Euclidean line $\R$ resolves ambiguity only for fixed sequences but our notion of origin is often relative.
\medskip

In practice, it is natural to consider the two sequences $\{0,1\}+3\Z$ and $\{0,2\}+3\Z$ equivalent because they both consist of pairs of points at a distance 1 translated with period 3.
Any translation in $\R$ is a 1-dimensional \emph{rigid motion} or orientation-preserving isometry.
\medskip

In general, an \emph{isometry} is any map that maintains all inter-point distances, hence includes all reflections $t\mapsto 2a-t$ around a fixed center $a\in\R$.
If we also allow uniform scaling, we get two more equivalence relations: \emph{similarity} (an affine map $t\mapsto at+b$ for $a\neq 0$ and any $b\in\R$) and \emph{orientation-preserving similarity} (an affine map $t\mapsto at+b$ for $a>0$).
\medskip

Now we introduce more general periodic point sets that model all solid crystalline materials (periodic crystals) whose structures are determined through diffraction patterns in a rigid form.
Hence the rigid motion is the strongest practical equivalence of periodic crystals.
% that remain unchanged under translations and rotations.

\begin{dfn}[a lattice $\La$ and a periodic point set $S$ in $\R^n$]
\label{dfn:periodic_set}
A \emph{lattice} $\La\subset\R^n$ is the infinite set $\{\sum\limits_{i=1}^n c_iv_i \mid c_i\in\Z\}$ of all integer linear combinations of a linear basis $v_1,\dots,v_n$ in $\R^n$.
A basis of $\La$ spans a parallelepiped $U(v_1,\dots,v_n)=\{\sum\limits_{i=1}^n c_iv_i \mid 0\leq c_i<1\}$ called a \emph{unit cell}.
For a basis $v_1,\dots,v_n$ of a lattice $\La$ and finite \emph{motif} of points $p_1,\dots,p_m\in U(v_1,\dots,v_n)$, the \emph{periodic point set} $S$ is the Minkowski sum $S=M+\La=\{u+v \mid u\in M, v\in\La\}\subset\R^n$.
\bs
\end{dfn}

For $n=1$, a periodic point set becomes a periodic sequence whose period interval $[0,l)$ is a unit cell.
A periodic point set can be visualized as a finite union of lattice images $\La+p$ obtained from a given lattice $\La\subset\R^n$ by shifting the origin to each motif point $p\in M\subset U$.
%\medskip

\begin{figure}[h]
\includegraphics[width=\textwidth]{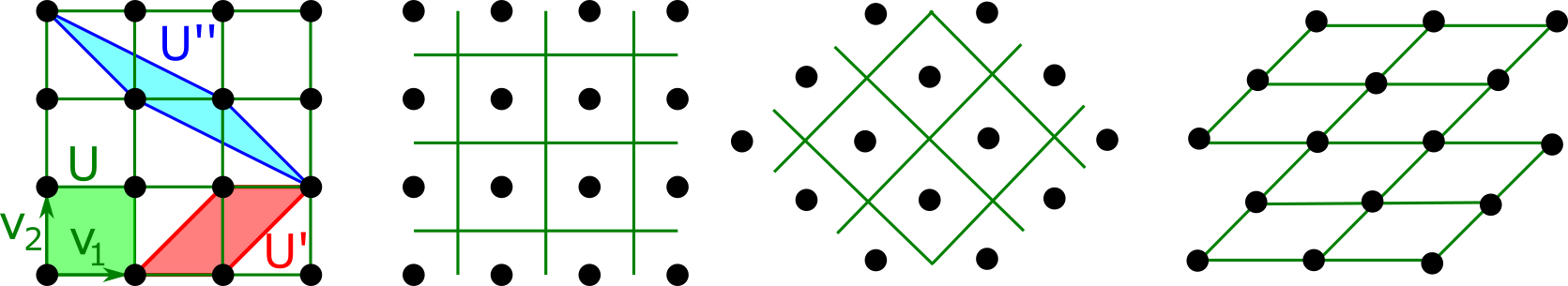}
\caption{The same square lattice (up to isometry) can be generated by infinitely many bases: in the left hand side picture, $v_1=(1,0),v_2=(0,1)$ span the green square unit cell $U$; $v_1=(1,0)$, $v_2=(1,1)$ span the red unit cell $U'$; $v_1=(-1,1),v_2=(2,-1)$ span the blue cell $U''$; and so on.}
\label{fig:square_lattice_cells}
\end{figure}

Any lattice can be generated by infinitely many bases, see Fig.~\ref{fig:square_lattice_cells}.
Even if we fix a basis, different motifs can lead to identical (up to translation) periodic point sets.
This ambiguity motivates us to study metrics on isometry classes of periodic point sets without a fixed basis.
%Fig.~\ref{fig:square_lattice_perturbations}~(right) shows that ever-present atomic vibrations can discontinuously affect all discrete invariants such as symmetry groups, and even the minimal cell doubles in volume.
%Hence a metric between periodic point sets should be continuous as now required in Problem~\ref{pro:metric}(e).

\begin{figure}[h]
\includegraphics[height=19mm]{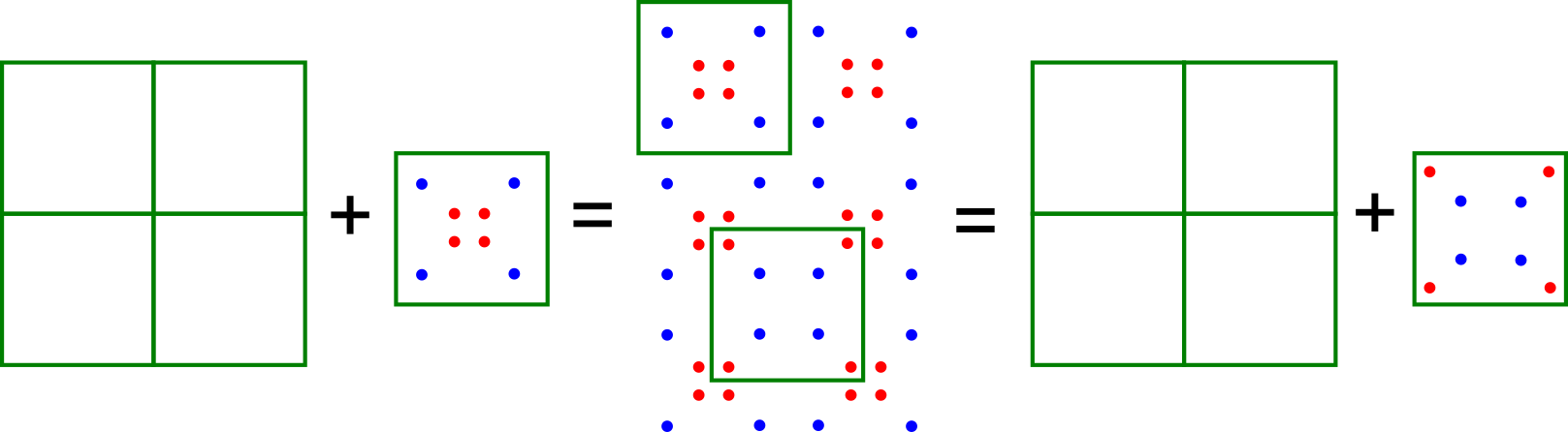}
\hspace*{2mm}
\includegraphics[height=19mm]{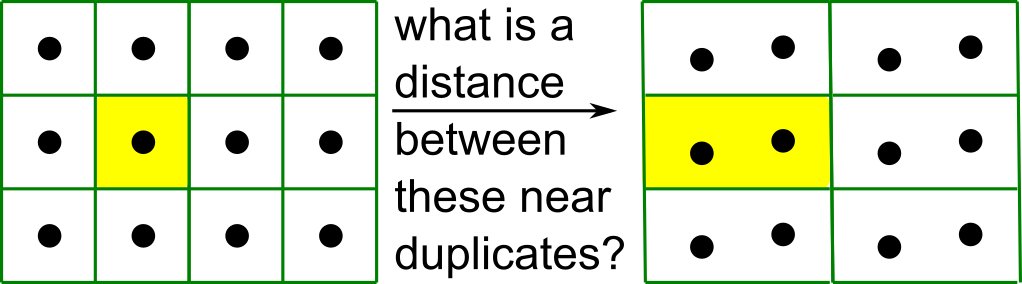}
\caption{\textbf{Left}: any periodic point set (up to isometry) can be given by many sums $\La+M$ in Definition~\ref{dfn:periodic_set}. 
\textbf{Right}: Problem~\ref{pro:metric}(e) asks for a continuous metric on nearly identical sets.}
\label{fig:square_lattice_perturbations}
\end{figure}

%Since the bottleneck distance $d_B$ involves the minimization over infinitely many bijections and points, $d_B$ is not computable in practice but can be efficiently used as an upper bound for computable metrics on periodic point sets, see condition~(e) in Problem~\ref{pro:metric} below.
Since a periodic point set $S$ with $m$ motif points in a unit cell has an input length of $O(m)$, we define the \emph{size} of $S$ as $|S|=m$.
Despite the recent progress in dimensions $n>1$ discussed in section~\ref{sec:review}, the following problem remained open even for dimension $n=1$.

\begin{pro}[continuous metric on isometry classes of periodic point sets]
\label{pro:metric}
Find a metric $d$ satisfying all metric axioms and the two practically important conditions (d,e) below.
\smallskip

\noindent
(a) \emph{first axiom} : 
$d(S,Q)=0$ if and only if periodic point sets $S\cong Q$ are isometric in $\R^n$;
\smallskip

\noindent
(b) \emph{symmetry axiom} : 
$d(S,Q)=d(Q,S)$ for any periodic point sets $S,Q\subset\R^n$;
\smallskip

\noindent
(c) \emph{triangle inequality} : 
$d(S,T)\leq d(S,Q)+d(Q,T)$ for any periodic point sets $S,Q,T\subset\R^n$;
\smallskip

\noindent
(d) \emph{computability} : $d(S,Q)$ can be exactly computed in a polynomial time in $\max\{|S|,|Q|\}$.
\smallskip

\noindent
(e) \emph{continuity} : 
let a periodic point set $Q$ be obtained from $S$ by perturbing each point of $S$ within its $\ep$-neighborhood,
then $d(S,Q)\leq C\ep$ for a constant $C$ and any such sets $S,Q$.
\bs
\end{pro}

Problem~\ref{pro:metric} was stated for isometry, which can be replaced by closely related equivalences such as rigid motion and similarity.
The first axiom in~\ref{pro:metric}(a) resolves the isometry problem.
To detect an isometry $S\cong Q$, it suffices to check if the distance vanishes: $d(S,Q)=0$.  
\medskip

The first three conditions in Problem~\ref{pro:metric}(a,b,c) imply the positivity $d\geq 0$ of any metric. 
\medskip

Continuity condition~\ref{pro:metric}(e) is non-trivial already in dimension $n=1$ because nearly identical periodic sequences might have very different periods. 
All conditions of Problem~\ref{pro:metric} were not satisfied by past work
%. Any satisfactory solution should work for the simplest case $n=1$, which remained open 
and will be fulfilled by Theorems~\ref{lem:Elm},~\ref{thm:complexity},~\ref{thm:continuity}.
\medskip

The proposed solution to Problem~\ref{pro:metric} for $n=1$ will be extended to higher dimensions for general 1-periodic sequences with values in $\R^{n-1}$, motivated by multivariate time series \cite{franses2004periodic}.

\begin{dfn}[high-dimensional 1-periodic sequences]
\label{dfn:high-dim_sequence}
Let $\vec e_1$ be the unit vector along the first coordinate axis in the product $\R\times\R^{n-1}$.
For a \emph{period} $l>0$, a \emph{motif} $M$ is a finite set of points $p_1,\dots,p_m$ in the slice $[0,l)\times\R^{n-1}$ of the width $l>0$.
We assume that the the times $t(p_1),\dots,t(p_m)$ under \emph{time projection} $t:[0,l)\times\R^{n-1}\to[0,l)$ are distinct, while the values $v(p_1),\dots,v(p_m)$ under the \emph{value projection} $v:[0,l)\times\R^{n-1}\to\R^{n-1}$ are arbitrary.
A \emph{high-dimensional 1-periodic sequence} $S=M+l\vec e_1\Z$ is the infinite sequence of points $p(i+mj)=p_i+jl\in\R^n$ indexed by $i+mj$, where $j\in\Z$ and $i=1,\dots,m$.
\bs
\end{dfn}

A 1-dimensional periodic sequence in Definition~\ref{dfn:periodic_sequence} is the simplest case $n=1$ of Definition~\ref{dfn:high-dim_sequence} when value projections are empty.
Section~\ref{sec:review} reviews past work related to Problem~\ref{pro:metric}.
%Main Theorems~\ref{thm:complexity} and~\ref{thm:continuity} solve Problem~\ref{pro:metric} for the first non-trivial case $n=1$.
Section~\ref{sec:metric} introduces elastic metrics and proves their computability and continuity fulfilling all conditions of Problem~\ref{pro:metric} for $n=1$.
Section~\ref{sec:finite} adapts the classical distance matrix of a finite sequence for a new metric convenient for cyclic shifts of points.
Section~\ref{sec:high-dim} extends the results above to complete invariants and continuous metrics for 1-periodic sequences in $\R^n$.
Section~\ref{sec:discussion} discusses the practical impact on protein biology and materials science.
 
%Further Theorems~\ref{thm:TVI} and~\ref{thm:high-Elm} solve the analog of Problem~\ref{pro:metric} for high-dimensional sequences $S\subset\R\times\R^{n-1}$ considered up to a product of isometries $f\times g$ of $\R\times\R^{n-1}$ so that $f$ translates the time projection $t(S)$ within $\R$ while $g$ is an isometry of $\R^{n-1}$ applied to the value projection $v(S)$ under $v:\R\times\R^{n-1}\to\R^{n-1}$.
%Example~\ref{6-point-pair}

%2========================
\section{A review of the past work on isometry classifications and metrics}
\label{sec:review}

\begin{figure}[h]
\centering
\includegraphics[height=15.5mm]{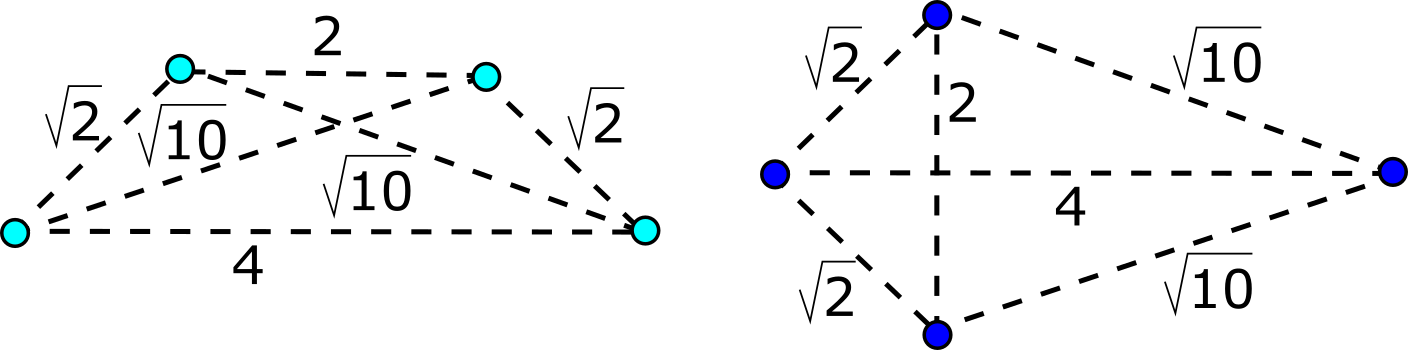}
\includegraphics[height=15.5mm]{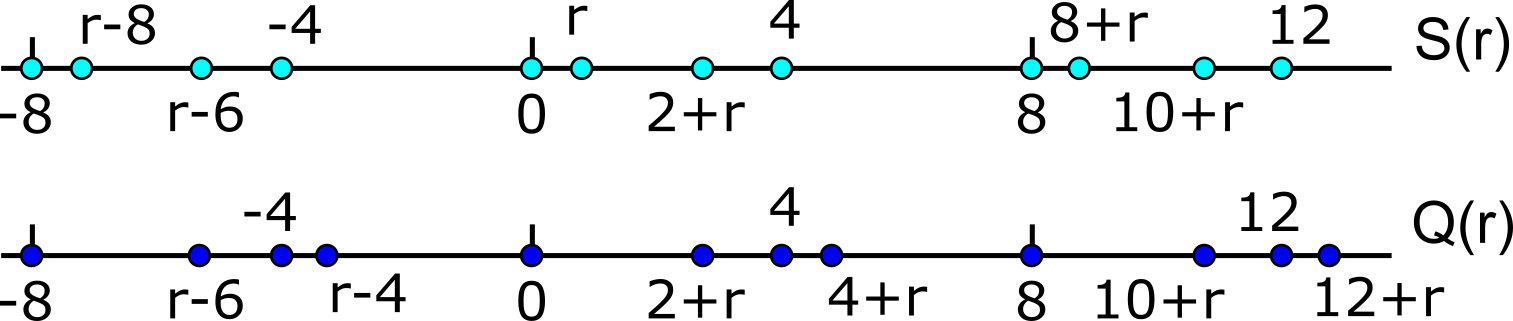}
\caption{\textbf{Left}: sets $K=\{(\pm 2,0),\; (\pm 1,1)\}$ and $T=\{(\pm 2,0),(-1,\pm 1)\}$ can not be distinguished by pairwise distances $\sqrt{2},\sqrt{2},2,\sqrt{10},\sqrt{10},4$.
\textbf{Right}: sequences $S(r)=\{0,r,2+r,4\}+8\Z$ and $Q(r)=\{0,2+r,4,4+r\}+8\Z$ for $0<r\leq 1$ have the same Patterson
function \cite[p.~197, Fig.~2]{patterson1944ambiguities}.}
\label{fig:non-isometric_pairs}
\end{figure}

For a finite set $S\subset\R^n$, the collection of all pairwise distances  is a complete invariant in general position \cite{boutin2004reconstructing} meaning that almost any finite set can be uniquely reconstructed up to isometry from the set of all pairwise distances.
The non-isometric 4-point sets in Fig.~\ref{fig:non-isometric_pairs} are a classical counter-example to the completeness of this distribution of all pairwise distances.
\medskip

The isometry classification of finite point sets was algorithmically resolved by \cite[Theorem~1]{alt1988congruence} saying that the existence of an isometry between two $m$-point sets in $\R^n$ can be checked in time $O(m^{n-2}\log m)$.
The algorithm from \cite{brass2000testing} checks if two finite sets of $m$ points are isometric in time $O(m^{\lceil n/3\rceil}\log m)$, so $O(m\log m)$ in $\R^3$ \cite{brass2004testing}.
The latest advance is the $O(m\log m)$ algorithm in $\R^4$ \cite{kim2016congruence}, see other significant results in \cite{ruggeri2008isometry,yang2015go, maron2016point,dym2019linearly}.
The Euclidean Distance Geometry
\cite{liberti2017euclidean} studies the related problem of uniquely embedding (up to isometry of $\R^n$) an abstract graph whose straight-line edges must have specified lengths.
M\'emoli's work on \emph{local distributions of distances }\cite{memoli2011gromov}, also known as \emph{shape distributions} \cite{osada2002shape, belongie2002shape, grigorescu2003distance, manay2006integral, pottmann2009integral}, for general metric spaces is closest to the new invariants of sequences.
\medskip

Patterson \cite{patterson1944ambiguities} was probably the first to systematically study periodic point sets.
He visualized any periodic sequence $S=\{p_1,\dots,p_m\}+l\Z\subset\R$ in a circle of a length $l$ but described its isometry classes by the complicated \emph{distance array} defined as the anti-symmetric $m\times m$ matrix of differences $p_i-p_j$ for $i,j\in\{1,\dots,m\}$.   
Gr{\"u}nbaum and Moore considered rational-valued periodic sequences given by complex numbers on the unit circle
and proved \cite[Theorem~4]{grunbaum1995use} that the combinations of $r$-factor products of complex numbers up to $r=6$ suffice to distinguish all such sequences up to translation.
This approach fixes a period of a sequence, hence not leading to a continuous metric even for $n=1$ in Problem~\ref{pro:metric}.
\medskip

Atomic vibrations are natural to measure by the maximum deviation of atoms from their initial positions as Problem~\ref{pro:metric}(e).
The maximum deviation can be small while taking a sum over infinitely many perturbed points is often infinite.
This deviation is defined as the bottleneck distance via  bijections between atoms, which can be displaced but cannot vanish.

\begin{dfn}[bottleneck distance $d_B$]
\label{dfn:bottleneck}
For any finite or periodic point sets $S,Q\subset\R^n$, the \emph{bottleneck distance} $d_B(S,Q)=\inf\limits_{h:S\to Q}\sup\limits_{p\in S}|p-h(p)|$ is minimized over all bijections $h:S\to Q$, where the Euclidean norm  of a vector $v=(x_1,\dots,x_m)\in\R^m$ is $|v|=\sqrt{\sum\limits_{i=1}^m x_i^2}$. 
\bs
\end{dfn}

\begin{exa}[infinite bottleneck]
\label{exa:dB_infinite}
We show that $S=\Z$ and $Q=(1+\de)\Z$ for any $\de>0$ have $d_B(S,Q)=+\infty$.
Assuming that $d_B(S,Q)$ is finite, consider an interval $[-N,N]\subset\R$ containing $2N+1$ points of $S$.
If there is a bijection $g:S\to Q$ such that $|p-g(p)|\leq d_B$ for all points $p\in S$, then the image of $2N+1$ points $S\cap [-N,N]$ under $g$ should be within the interval $[-N-d_B,N+d_B]$.
The last interval contains only $1+\frac{2(N+d_B)}{1+\de}$ points, which is smaller than $1+2N$ when $\frac{N+d_B}{1+\de}<N$, $d_B<\de N$, which is a contradiction for $N>\frac{d_B}{\de}$.
\bs
\end{exa}

If we consider only periodic point sets $S,Q\subset\R^n$ with the same density (or unit cells of the same size), the bottleneck distance $d_B(S,Q)=\inf\limits_{g:S\to Q}\; \sup\limits_{a\in S}|a-g(a)|$ takes finite values and becomes a well-defined wobbling distance \cite{carstens1999geometrical}, which is unfortunately discontinuous.

\begin{exa}[discontinuous bottleneck]
\label{exa:dB_discontinuous}
Slightly perturb the basis $(1,0),(0,1)$ of the integer lattice $\Z^2$ to the basis vectors $(1,0),(\ep,1)$ of the new lattice $\La$.
We prove that $d_B(\La,\Z^2)\geq\frac{1}{2}$ for any $\ep>0$.
Map $\R^2$ by $\Z^2$-translations to the unit square $[0,1]^2$ with identified opposite sides (a torus).
Then $\Z^2$ maps to one point represented by the corners of the square $[0,1]^2$.
The perturbed lattice $\La$ maps to the sequence of points $\{k\ep\pmod{1}\}_{k=0}^{+\infty}\times\{0,1\}$ in the horizontal edges.
If $d_B(\La,\Z^2)=r<\frac{1}{2}$, then all above points should be covered by the closed disks of the radius $r$ centered at the corners of $[0,1]^2$.
For $0<\ep<\frac{1}{2}-r$, we can find a point $k\ep$ that is between $r,1-r$, hence not covered by these disks, so $d_B(\La,\Z_2)\geq\frac{1}{2}$.
\bs 
\end{exa}

The periodic sequences $S(r),Q(r)$ in Fig.~\ref{fig:non-isometric_pairs} emerged as first examples with identical infinite distributions of distances (or diffraction patterns).
They are distinguished by recent Pointwise Distance Distributions  \cite{widdowson2021pointwise} but not by the simpler Average Minimum Distances \cite{widdowson2022average}.
However, the latter invariants distinguish the even more interesting periodic sequences $S_{15} = \{0,1,3,4,5,7,9,10,12\}+15\Z$ and $Q_{15} = \{0,1,3,4,6,8,9,12,14\}+15\Z$.
These sets have identical density functions \cite[Example~11]{anosova2021introduction}, which form an infinite sequence of invariant functions \cite{edels2021} depending on a variable radius.
The density functions were fully described for periodic sequences by \cite[Theorems~5, 7]{anosova2022density} but didn't lead to a metric due to incompleteness. 
\medskip

Appendix~\ref{sec:isoset} reviews the recent isoset invariant for any periodic point sets in $\R^n$.
This isoset reduces the isometry classification of any periodic point sets in $\R^n$ to the simpler equivalence for finite sets up to rotations around a fixed center.
The continuous metric on isosets \cite[section~7]{anosova2021introduction} requires minimizations over infinitely many rotations for any dimension $n\geq 2$.
For all dimensions $n\geq 1$, this metric also depends on a stable radius $\al$.
The above disadvantages left Problem~\ref{pro:metric} open even for periodic sequences in dimension $n=1$.
%whose easy upper bound can be maximized for comparing real periodic structures in any finite dataset.

%3========================
\section{Elastic metrics on spaces of isometry classes of periodic sequences}
\label{sec:metric}

This section introduces a simple distance-based invariant, which can be expanded to both Patterson's distance array and isoset.
The completeness of these smallest distance lists ($\SDL$) is proved up to four equivalences in Proposition~\ref{prop:SDL}.
The hardest challenge will be to introduce a metric on these $\SDL$ invariants to satisfy all conditions of Problem~\ref{pro:metric}.

\begin{dfn}[smallest distance lists $\SDL^o$ and $\SDL$] 
\label{dfn:SDL}
For a periodic sequence $S=\{p_1,\dots,p_m\}+l\Z\subset\R$,
let $d_i=p_{i+1}-p_{i}$ be the distance between successive points of $S$ for $i=1,\dots,m$, where $p_{m+1}=p_1+l$.
Consider the lexicographic order on ordered lists so that $(d_1,\dots,d_m)<(d'_1,\dots,d'_m)$ if $d_1=d'_1,\dots,d_i=d'_i$ for some $0\leq i<m$, where $i=0$ means the empty set of identities, and $d_{i+1}<d'_{i+1}$.
The oriented \emph{smallest distance list} $\SDL^o(S)$ is the lexicographically smallest list obtained from $(d_1,\dots,d_m)$ by cyclic permutations. 
The unoriented \emph{smallest distance list} $\SDL(S)$ is the lexicographically smallest list obtained from $(d_1,d_2,\dots,d_m)$ and the reversed list $(d_m,d_{m-1},\dots,d_1)$ by cyclic permutations.
\bs
\end{dfn}

The periodic sequences $S_2=\{0,1\}+3\Z$ and $3-S_2=\{0,2\}+3\Z$ have the same $\SDL^o=(1,2)=\SDL$.
The periodic sequences $S_3=\{0,1,3\}+6\Z$ and  $6-S_3=\{0,3,5\}+6\Z$ have $\SDL(S_3)=(1,2,3)=\SDL(6-S_3)$ and $\SDL^o(S_3)=(1,2,3)\neq (1,3,2)=\SDL^o(6-S_3)$.

\begin{prop}[classifications of periodic sequences]
\label{prop:SDL}
For any periodic sequence $S=\{p_1,\dots,p_m\}+l\Z\subset\R$ with a minimum period $l>0$, the smallest distance lists $\SDL^o(S)$, $\SDL(S)$ are complete invariants up to translation and isometry, respectively.
If we scale $\SDL(S)$, $\SDL^o(S)$ so that the sum of distances is 1, the \emph{normalized distance lists} $\NDL(S)$, $\NDL^o(S)$ are complete up to similarity and orientation-preserving similarity, respectively.
\bs
\end{prop}
\begin{proof}[Proof of Proposition~\ref{prop:SDL}]
Any translation in $\R$ preserves the cyclic order of the points $p_i$ and their inter-point distances $d_i=p_{i+1}-p_i$.
Up to translation in $\R$, the ordered distance list $(d_1,\dots,d_m)$ can change only by cyclic permutation, so $\SDL(S)$ is invariant.
Any reflection preserves all distances but reverses their order to $(d_m,d_{m-1},\dots,d_1)$ for a fixed orientation of $\R$. 
The completeness of $\SDL^o(S)$ and $\SDL(S)$ follows by reconstructing the points $p_i=\sum\limits_{j=1}^{i-1}d_j$, where the first point $p_1=0$ is at the origin, up to translation and isometry, respectively.  
\medskip

Finally, normalizing the distance lists $\SDL^o(S)$ and $\SDL(S)$ to make the sum of distances equal to 1 is equivalent to scaling a periodic sequence to make its period equal to 1.
\end{proof}

\begin{figure}[h]
\centering
\includegraphics[height=22mm]{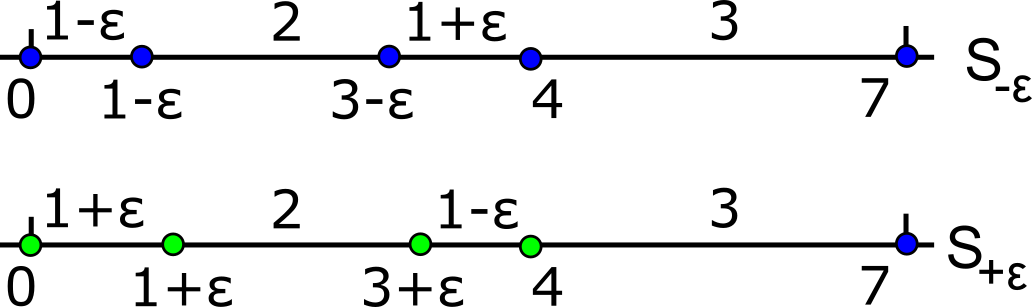}
\hspace*{6mm}
\includegraphics[height=22mm]{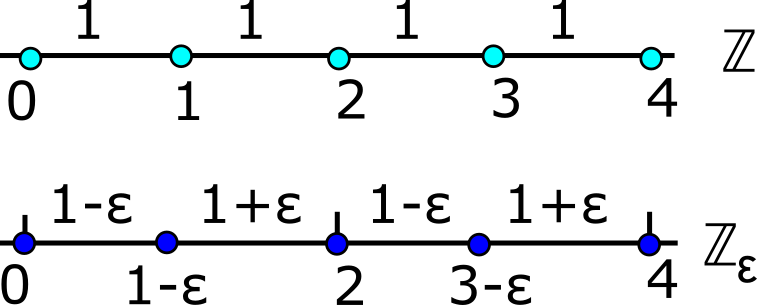}
\caption{\textbf{Left}: the distance lists $\SDL^o$ of the nearly identical sequences
$S_{\pm\ep}=\{0,1\pm\ep,3\pm\ep,4\}+7\Z$ are not close, see Example~\ref{exa:SDL}(a), which motivates a metric minimized over cyclic permutations. 
\textbf{Right}: $\Z$ and its perturbation $\Z_\ep$ have incomparable $\SDL(\Z)=(1)$ and $\SDL(\Z_\ep)=(1-\ep,1+\ep)$, which motivates elastic metrics in Definition~\ref{dfn:Elm} based on multiples of sequences in Definition~\ref{dfn:multiple}.
}
\label{fig:SDL}
\end{figure}

The naive subtraction of SDLs reveals the important discontinuity of the component-wise comparison illustrated in Example~\ref{exa:SDL} and Fig.~\ref{fig:SDL}, which will be resolved %by new elastic metrics 
in Definition~\ref{dfn:Elm}.

\begin{exa}%[discontinuity]
\label{exa:SDL} 
%\textbf{(a)}
The periodic sequence $S_0=\{0,1,3,4\}+7\Z$ has perturbations $S_{\pm\ep}=\{0,1\pm\ep,3\pm\ep,4\}+7\Z$ for any small $\ep>0$.
The distance lists $\SDL^o(S_{-\ep})=(1-\ep,2,1+\ep,3)$ and $\SDL^o(S_{+\ep})=(1-\ep,3,1+\ep,2)$ have the large component-wise difference $(0,1,0,-1)$ because the minimum distance $1-\ep$ is followed by different distances $2<3$ in the nearly identical $S_{\pm\ep}=\{0,1\pm\ep,3\pm\ep,4\}+7\Z$ for any $\ep>0$, see Fig.~\ref{fig:SDL}~(left).
This discontinuity will be resolved by minimizing over cyclic permutations but there is one more obstacle below.
\bs
\end{exa}

It seems natural to always reduce a period of $S=\{p_1,\dots,p_m\}+l\Z\subset\R$ to a minimum positive value $l>0$.
The list $\SDL(S)=(d_1,\dots,d_m)$ of a fixed size $m$ cannot be directly used for comparing sequences that have different numbers of motif points, see Fig.~\ref{fig:SDL}~(right).

\begin{dfn}[multiple $kS$ of a periodic sequence $S$]
\label{dfn:multiple}
For any integer $k>1$ and a periodic sequence $S=\{p_1,\dots,p_m\}+l\Z\subset\R$, define the \emph{multiple} periodic sequence as $kS=\{p_1+jl,\dots,p_m+jl\}_{j=0}^{k-1}+kl\Z$ with $km$ motif points in the period interval $[0,kl)$.
\bs
\end{dfn}

Definition~\ref{dfn:Elm} will introduce two elastic metrics comparing any periodic sequences $S,Q$ through their multiples from Definition~\ref{dfn:multiple} with the same number $m$ of motif points.
\medskip

For any ordered list $(d_1,\dots,d_m)$, let $C^o(m)$ be the \emph{cyclic} (or dihedral) group of $m$ \emph{cyclic} permutations that are iterations of $(d_1,d_2,\dots,d_m)\mapsto(d_2,\dots,d_m,d_1)$.
Let $C(m)\supset C^o(m)$ denote the larger group of $2m$ permutations also including the reversed cyclic permutations composed of the reversion $(d_1,d_2,\dots,d_m)\mapsto(d_m,\dots,d_2,d_1)$ with all $\si\in C^o(m)$.
Denote the action of a permutation $\si$ on a vector $v=(d_1,d_2,\dots,d_m)$ by $\si(v)=(d_{\si(1)},\dots,d_{\si(m)})$.

\begin{dfn}[elastic metrics $\Elm^o$ and $\Elm$]
\label{dfn:Elm}
For any periodic sequences $S,Q\subset\R$, let $m=\lcm(|S|,|Q|)$ be the lowest common multiple of their motif sizes.
Consider the multiple sequences $\frac{m}{|S|}S$ and $\frac{m}{|Q|}Q$ from Definition~\ref{dfn:multiple}, which have the same number $m$ of motif points within the period intervals extended by the integer factors $\frac{m}{|S|}$ and $\frac{m}{|Q|}$, respectively.
\medskip
 
%\noindent 
The \emph{oriented elastic metric}
$\Elm^o(S,Q)=\min\limits_{\si\in C^o(m)}||\SDL^o(\frac{m}{|S|}S)-\si(\SDL^o(\frac{m}{|Q|}Q))||_{\infty}$ is minimized over all $m$ cyclic permutations $\si\in C^o(m)$ of $m$ ordered distances.
\medskip
 
%\noindent 
The (unoriented) \emph{elastic metric} $\Elm(S,Q)=\min\limits_{\si\in C(m)}||\SDL(\frac{m}{|S|}S)-\si(\SDL(\frac{m}{|Q|}Q))||_{\infty}$ is minimized over all $2m$ cyclic permutations $\si$ from the group $C(m)$.
\bs
\end{dfn}

\begin{exa}
\label{exa:Elm}
\textbf{(a)}
The sequences $S_3=\{0,1,3\}+6\Z$ and $6-S_3=\{0,3,5\}+6\Z$ have the same number $m=3$ of motif points.
Since $\SDL(S_3)=(1,2,3)=\SDL(6-S_3)$, the unoriented distance is $\Elm(S_3,6-S_3)=0$.
Indeed, $S_3$ and $6-S_3$ are related by reflection, which also follows from the first axiom in Lemma~\ref{lem:Elm}.
Since $\SDL^o(S_3)=(1,2,3)\neq (1,3,2)=\SDL^o(6-S_3)$, the oriented distance is $\Elm^o(S_3,6-S_3)=||(1,2,3)-(1,3,2)||_{\infty}=1$, which is the minimum over all cyclic permutations of $(1,3,2)$, e.g. $||(1,2,3)-(3,2,1)||_{\infty}=2$.
\medskip

\noindent
\textbf{(b)}
To compute the oriented distance between the periodic sequences $S_2=\{0,1\}+3\Z$ and $S_3=\{0,1,3\}+6\Z$, we consider
$3S_2=\{0,1,3,4,6,7\}+9\Z$ and $2S_3=\{0,1,3,6,7,9\}+12\Z$.
Then $\SDL^o(3S_2)=(1,2,1,2,1,2)$ and $\SDL^o(2S_3)=(1,2,3,1,2,3)$.
The component-wise comparison gives $\Elm^o(S_2,S_3)=||(1,2,1,2,1,2)-(1,2,3,1,2,3)||_{\infty}=2$ because any shifts of one sequence relative to the other leads to the maximum distance $|1-3|=2$.
\bs
\end{exa}

%4========================
%\section{Computability and continuity of elastic metrics under perturbations}
%\label{sec:properties}

Definition~\ref{dfn:Elm} used any (not necessarily minimal) periods of sequences $S,Q$ to define the elastic metrics.
Lemma~\ref{lem:multiples} shows that $\Elm^o(S,Q)$ and $\Elm(S,Q)$ are independent of a period and motivates the term \emph{elastic} for metrics on sequences with extendable periods.

\begin{lem}[elastic metrics for multiples $kS$]
\label{lem:multiples}
For any periodic sequences $S,Q\subset\R$, the functions $\Elm^o(kS,Q),\Elm(kS,Q)$ from Definition~\ref{dfn:Elm} are independent of a factor $k\geq 1$.
\bs
\end{lem}

The proof of Lemma~\ref{lem:multiples} used the fact the Minkowski norm $||v||_{\infty}$ remains unchanged when any vector $v=(x_1,\dots,x_m)$ is concatenated with copies of $v$.
Other Minkowski norms $||v||_q=(\sum\limits_{i=1}^m |x_i|^q)^{1/q}$ are not invariant under this transformation for $q\in[1,+\infty)$.
 
\begin{lem}[axioms for elastic metrics]
\label{lem:Elm}
$\Elm^o,\Elm$ from Definition~\ref{dfn:Elm} satisfy the metric axioms for any periodic sequences up to translation and isometry in $\R$, respectively.
\bs
\end{lem}

For the elastic metrics on periodic sequences in $\R$, Lemma~\ref{lem:Elm} fulfilled conditions~(a,b,c) of Problem~\ref{pro:metric}.
The remaining conditions (d,e) hold due to Theorems~\ref{thm:complexity} and~\ref{thm:continuity}. 
\medskip

Any periodic sequence $S=\{p_1,\dots,p_m\}+l\Z$ can be given by its $m$ motif points and period $l$.
Hence the input size for all time complexities below is $|S|=m$.

\begin{thm}[time complexity of elastic metrics]
\label{thm:complexity}
For any periodic sequences $S,Q\subset\R$, let $m=\lcm(|S|,|Q|)$ be the lowest common multiple of their motif sizes.
The elastic metrics $\Elm^o(S,Q)$ and $\Elm(S,Q)$ from Definition~\ref{dfn:Elm} can be computed in time $O(m^2)$.
\bs
\end{thm}

%Examples~\ref{exa:dB_infinite} and~\ref{exa:dB_discontinuous} show that the bottleneck distance $d_B(S,Q)$ can be infinite or discontinuous even if $S,Q$ are nearly identical lattices.
%Then Theorem~\ref{thm:continuity} will prove that the elastic metrics are continuous under perturbations of any periodic sequences.
%Theorem~\ref{thm:continuity} will prove a stronger upper bound in terms of the well-defined continuous metric on the complete isoset invariant isosets \cite[sections~6-7]{anosova2021introduction}.

Any 1-dimensional lattice $l\Z$ has $\SDL^o(l\Z)=(l)$ consisting of a single distance $l$.
Then any lattices $l\Z$ and $l'\Z$ have the elastic metrics $\Elm(l\Z,l'\Z)=|l-l'|=\Elm^o(l\Z,l'\Z)$.
In particular, $\Z$ and $(1+\de)\Z$ in Example~\ref{exa:dB_infinite} have the small distance $\de$ in the elastic metrics.

\begin{thm}[continuity of elastic metrics]
\label{thm:continuity}
The elastic metrics from Definition~\ref{dfn:Elm} satisfy the continuity $\Elm(S,Q)\leq\Elm^o(S,Q)\leq 2d_B(S,Q)$ for any periodic sequences $S,Q\subset\R$.
% with a minimum inter-point distance more than $2d_B(S,Q)$.
\bs
\end{thm}

%4========================
\section{Metrics on isometry classes of high-dimensional finite sequences}
\label{sec:finite}

This section studies two families of continuous and easy computable metrics on isometry classes of finite sequences of labeled (or indexed) points in $\R^n$.
The metrics are based on the classical distance matrix $\DM$ and the new cyclic distance matrix $\CDM$ in Definition~\ref{dfn:CDM}.
The latter matrix will be used in section~\ref{sec:high-dim} for elastic metrics on 1-periodic sequences in $\R^n$.
An isometry of $\R^n$ should preserve the order of points in such a sequence $p_1,\dots,p_m\in\R^n$.

\begin{dfn}[distance matrices $\DM$ and $\CDM$]
\label{dfn:CDM}
Let $T=\{p_1,\dots,p_m\}$ be an ordered sequence of $m$ points in $\R^{n}$.
In the \emph{distance matrix} $\DM(T)$ of the size $m\times m$, each element $\DM_{ij}(T)$ is the Euclidean distance $|p_j-p_{j}|$ for $i,j\in\{1,\dots,m\}$, so $d_{ii}=0$ for $i=1,\dots,m$.
\smallskip

In the \emph{cyclic distance matrix} $\CDM(T)$ of the size $(m-1)\times m$, each element $\CDM_{ij}(T)$ is the Euclidean distance $|p_j-p_{i+j}|$ for $i\in\{1,\dots,m-1\}$ and $j\in\{1,\dots,m\}$, where all indices are considered modulo $m$, for example, $p_{m+1}=p_1$.
Any cyclic permutation $\si\in C^o(m)$ of indices acts on the cyclic distance matrix $\CDM(Q)$ by cyclically shifting its $m$ columns.
\bs 
\end{dfn}

Any $m=3$ points in $\R^n$ with pairwise distances $d_{ij}$ have the distance matrix
$\DM=\left(\begin{array}{ccc}
0 & d_{12} & d_{13} \\
d_{12} & 0 & d_{23} \\
d_{13} & d_{23} & 0 
\end{array} \right)$ and 
the cyclic distance matrix
$\CDM=\left(\begin{array}{ccc}
d_{12} & d_{23} & d_{13} \\
d_{13} & d_{12} & d_{23} 
\end{array} \right)$.
$\CDM(T)$ is obtained from $\DM(T)$ by removing the zero diagonal and cyclically shifting each column so that the first row of $\CDM(T)$ has distances from $p_i$ to the next point $p_{i+1}$ in $T$.

\begin{figure}[h]
\centering
\includegraphics[width=\textwidth]{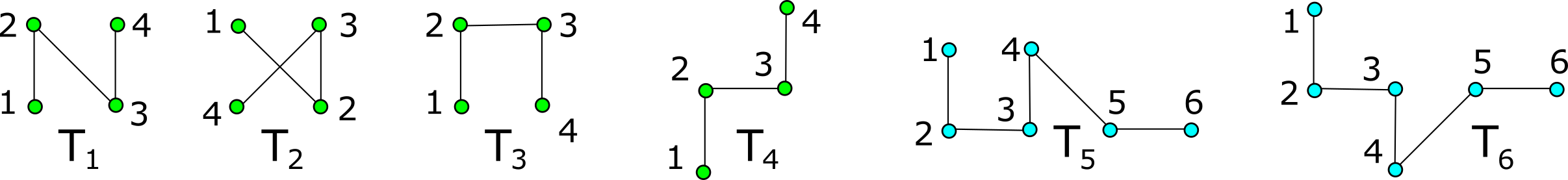}
\caption{These sequences are distinguished by their cyclic distance matrices in Example~\ref{exa:CDM}.}
\label{fig:finite_sequences}
\end{figure}

\begin{exa}[cyclic distance matrices]
\label{exa:CDM}
Fig.~\ref{fig:finite_sequences} shows the sequences $T_1,\dots,T_6\subset\R^2$ whose points are in the integer lattice $\Z^2$ so that the minimum inter-point distance is 1.
In each sequence, the points are connected by lines in the increasing order of their distances.
$\CDM(T_1)=\left(\begin{array}{cccc}
1 & \sqrt{2} & 1 & \sqrt{2} \\
1 & 1 & 1 & 1\\
\sqrt{2} & 1 & \sqrt{2} & 1
\end{array} \right)$,
$\CDM(T_2)=\left(\begin{array}{cccc}
\sqrt{2} & 1 & \sqrt{2} & 1 \\
1 & 1 & 1 & 1\\
1 & \sqrt{2} & 1 & \sqrt{2}
\end{array} \right)$ are different but related by a cyclic shift of columns. 
This shift of indices in $T_1$ gives a sequence isometric to $T_2$. 
Then
$\CDM(T_3)=\left(\begin{array}{cccc}
1 & 1 & 1 & 1 \\
\sqrt{2} & \sqrt{2} & \sqrt{2} & \sqrt{2}\\
1 & 1 & 1 & 1
\end{array} \right)$,
$\CDM(T_4)=\left(\begin{array}{cccc}
1 & 1 & 1 &  \sqrt{5}\\
\sqrt{2} & \sqrt{2} & \sqrt{2} & \sqrt{2}\\
\sqrt{5} & 1 & 1 & 1
\end{array} \right)$.
The CDMs of the sets $T_5,T_6$ differ only by distances $|p_1-p_4|=1$ in $T_5$ and $|p_1-p_4|=\sqrt{5}$ in the highlighted cells below.
If reduce the number $m-1$ of rows in $\CDM$ to the dimension $n=2$, the smaller matrices fail to distinguish the non-isometric sequences $T_5\not\cong T_6$.
%Similar examples can be constructed for  

\noindent
$T_5:\; \left(\begin{array}{cccccc}
1 & 1 & 1 &  \sqrt{2} & 1 & \sqrt{10}\\
\sqrt{2} & \sqrt{2} & 1 & \sqrt{5} & \sqrt{5} & 3 \\
\mc{yellow}{1} & 2 & 2 & \mc{yellow}{1} & 2 & 2 \\
\sqrt{5} & 3 & \sqrt{2} & \sqrt{2} & 1 & \sqrt{5} \\
\sqrt{10} & 1 & 1 & 1 & \sqrt{2} & 1
\end{array} \right)$,
$T_6:\; \left(\begin{array}{cccccc}
1 & 1 & 1 &  \sqrt{2} & 1 & \sqrt{10}\\
\sqrt{2} & \sqrt{2} & 1 & \sqrt{5} & \sqrt{5} & 3 \\
\mc{yellow}{\sqrt{5}} & 2 & 2 & \mc{yellow}{\sqrt{5}} & 2 & 2 \\
\sqrt{5} & 3 & \sqrt{2} & \sqrt{2} & 1 & \sqrt{5} \\
\sqrt{10} & 1 & 1 & 1 & \sqrt{2} & 1
\end{array} \right)$.
%\bs
\end{exa}

Recall that, for any matrix $k\times m$ re-written row-by-row as a vector $v\in\R^{km}$, the Minkowski norm $||v||_q=\left(\sum\limits_{i=1}^{km}|v_i|^q\right)^{1/q}$ in the limit case $q=+\infty$ is $||v||_{\infty}=\max\limits_{i=1}^{km}|v_i|^q$.
%For any $n\geq 1$, $\CDM$ has $\frac{n(n-1)}{2}$ repeated distances but has cyclically permutable columns.
%Proposition~\ref{prop:MCM} can be folklore but we couldn't find a reference.

\begin{prop}[metric $\MCM_q$ based on cyclic distance matrices]
\label{prop:MCM}
For any Minkowski metric with a parameter $q\in[1,+\infty]$ and ordered sequences $S,T\subset\R^{n-1}$ of $m$ points,
the function $\MCM_q(S,T)=||\DM(S)-\DM(T)||_q=||\CDM(S)-\CDM(T)||_q$ defines a metric on equivalence classes of $S,T$ considered up to isometry of $\R^n$ preserving the order of points.
Then $\DM(T)$ and $\CDM(T)$ are complete isometry invariants of an ordered sequence $T$.
The matrices $\DM(T),\CDM(T)$ and the metric $\MCM_q(S,T)$ are computed in time $O(m^2)$.
\bs
\end{prop}

\begin{exa}[metric $\MCM_{q}$]
\label{exa:MCM}
For any $q\in[1,+\infty)$, we use cyclic distance matrices from Example~\ref{exa:CDM}
to compute
$\MCM_q(T_3,T_4)=(\sqrt{5}-1)2^{1/q}$,
$\MCM_q(T_1,T_3)=(\sqrt{2}-1)8^{1/q}$, and
$\MCM_q(T_1,T_4)=(6(\sqrt{2}-1)^q+2(\sqrt{5}-\sqrt{2})^q)^{1/q}$.
The triangle inequality holds:
$$\MCM_q(T_1,T_3)+ \MCM_q(T_1,T_3)\geq (6(\sqrt{2}-1)^q)^{1/q}+(\sqrt{2}-1)2^{1/q}\geq\MCM_q(T_1,T_3)$$
 due to $6^{1/q}+2^{1/q}\geq 8^{1/q}$, which follows by taking both sides to the power $q$.
In the limit case $q=+\infty$, the above inequality becomes $(\sqrt{5}-\sqrt{2})+(\sqrt{2}-1)=\sqrt{5}-1$.
The final non-isometric sequences $T_5\cong T_6$ have
$\MCM_q(T_5,T_6)=(\sqrt{5}-1)2^{1/q}$.
\bs
\end{exa}

%5========================
\section{Metrics on isometry classes of high-dimensional periodic sequences}
\label{sec:high-dim}

This section solves the analog of Problem~\ref{pro:metric} for high-dimensional periodic sequences $S\subset\R\times\R^{n-1}$ from Definition~\ref{dfn:high-dim_sequence}.
Such a sequence models a multivariate time series \cite{bale2007kaleidomaps} with values in $\R^{n-1}$.
We focus on the natural equivalence $f\times g$, where $f$ is a translation in $\R$ and $g$ is any isometry in $\R^{n-1}$.
All results can be adapted to other cases, for example, including reflections in $\R$ or restricting to orientation-preserving isometry in $\R^{n-1}$.
%\medskip

\begin{dfn}[time-value invariant $\TVI$]
\label{dfn:TVI}
Let $S=\{p_1,\dots,p_m\}+l\vec e_1\Z$ be any high-dimensional periodic sequence in $\R\times\R^{n-1}$. 
The \emph{time projection} $t:\R\times\R^{n-1}\to\R$ gives the 1-dimensional periodic sequence $t(S)$ with the \emph{distance list} $\DL^o(t(S))=(d_1,\dots,d_m)$ of $d_i=t(p_{i+1})-t(p_i)$ ordered as motif points for $i=1,\dots,m$, where $t(p_{m+1})=t(p_1)+l$. 
The \emph{value projection} $v:\R\times\R^{n-1}\to\R^{n-1}$ gives the ordered \emph{value sequence} $v(S)$ of $m$ points $v(p_1),\dots,v(p_m)\in\R^{n-1}$. 
The time-value invariant is the pair $\TVI(S)=(\DL^o(t(S)),\CDM(v(S)))$ considered up to cyclic permutations $\si\in C^o(m)$ simultaneously acting on the distance list $\DL^o(t(S))$ and the cyclic distance matrix $\CDM(v(S))$.
\bs
\end{dfn}
 
The equivalence of high-dimensional sequences is considered 
 the sense of Definition~\ref{dfn:high-dim_sequence}.
 
\begin{thm}[classification of high-dimensional sequences]
\label{thm:TVI}
Any high-dimensional periodic sequences $S,Q\subset\R\times\R^{n-1}$  with $m$ motif points are equivalent in the sense of Definition~\ref{dfn:high-dim_sequence} if and only if there is a cyclic permutation $\si\in C^o(m)$ such that $\si(\TVI(S))=\TVI(Q)$.
\bs
\end{thm}

For any high-dimensional periodic sequence $S=\{p_1,\dots,p_m\}+l\vec e_1\Z$ in $\R\times\R^{n-1}$ and integer $k\geq 2$, its \emph{multiple} is $kS=\{p_1+jl\vec e_1,\dots,p_m+jl\vec e_1\}_{j=0}^{k-1}+kl\vec e_1\Z$. % similar to Definition~\ref{dfn:multiple}. 
For any matrix $M$, the Minkowski norm $||M||_{\infty}$ is the maximum absolute value of its real elements.
\medskip

Definition~\ref{dfn:high-Elm} extends the elastic metric $\Elm^o$ to high-dimensional periodic sequences by taking the maximum with the Minkowski metric on cyclic distance matrices $\CDM$ of value projections.
Instead of the maximum, one can combine two metrics in many other ways.

\begin{dfn}[elastic metric for high-dimensional sequences]
\label{dfn:high-Elm}
For any generic periodic sequences $S,Q\subset\R\times\R^{n-1}$, let $m=\lcm(|S|,|Q|)$ be the lowest common multiple of their motif sizes.
Consider the multiple sequences $\frac{m}{|S|}S$ and $\frac{m}{|Q|}Q$, which have the same number $m$ of motif points.
The \emph{oriented elastic metric}
$\Elm^o(S,Q)=\min\limits_{\si\in C^o(m)}\max\{d_t,d_v\}$ is minimized over all $m$ permutations $\si\in C^o(m)$, where
$d_t=||\DL^o(\frac{m}{|S|}S)-\si(\DL^o(\frac{m}{|Q|}Q))||_{\infty}$,
$d_v=||\CDM(v(\frac{m}{|S|}S))-\si(\CDM(v(\frac{m}{|Q|}Q)))||_{\infty}$, $\si$ acts on $\CDM$ by cyclic shifts of columns.
\bs
\end{dfn}

\begin{comment}
Example~\ref{exa:3-point_pair} shows that the condition on minimum distances in Theorem~\ref{thm:high-Elm} is essential.

\begin{exa}[harder periodic sequences]
\label{exa:3-point_pair}
For any $\ep>0$, consider the periodic sequences $Q^{\pm}=M^{\pm}+2\Z\vec e_1\subset\R\times\R$ with a period slice of width 2 and 3-point motifs $M^+=\{(0,0),(1-\ep,2),(1+\ep,-1)\}$ and $M^+=\{(0,0),(1-\ep,-1),(1+\ep,2)\}$.
Then $d_B(Q^-,Q^+)=2\ep$ and Theorem~\ref{thm:high-Elm} applies only for perturbations that don't swap the time projections $1\pm\ep$.
\medskip

We show that $\Elm^o(Q^-,Q^+)=1-3\ep$ as follows.
The time projections $t(Q^\pm)=(0,1-\ep,1+\ep)$ have the same smallest distance list $\SDL=(2\ep,1-\ep,1-\ep)$.
In Definition~\ref{dfn:high-Elm}, we have $m=3=|Q^\pm|$, so $d_t=||\DL(Q^+)-\DL(Q^-)||_{\infty}$ and $d_v=||\CMD(Q^+)-\CDM(Q^-)||_{\infty}$, where $\si$ is one of three shifts in $C^o(3)$
The trivial shift $\si$ gives $d_t=0$ 
\bs
\end{exa}
\end{comment}

\begin{exa}[hardest periodic sequences]
\label{exa:6-point_pair}
The latest counter-example \cite[Fig.~3]{pozdnyakov2022incompleteness} to the completeness of past distance-based invariants is the pair of the sequences $A^{\pm}\subset\R\times\R^2$ with a period $p$ and 6 motif points: %described in Fig.~\ref{fig:6-point_periodic_sequences}.
$A^+=\{W',C_+,V,W,C'_+,V'\}+p\vec e_1\Z$ and  $A^-=\{W',C_-,V,W,C'_-,V'\}+p\vec e_1\Z$, where $V=(v_x,v_y,0)$,  $W=(\frac{p}{2},w_y,w_z)$, $C_{\pm}=(\frac{p}{4},c_y,\pm c_z)$.
%\medskip

\begin{figure}[h]
\centering
\includegraphics[width=\textwidth]{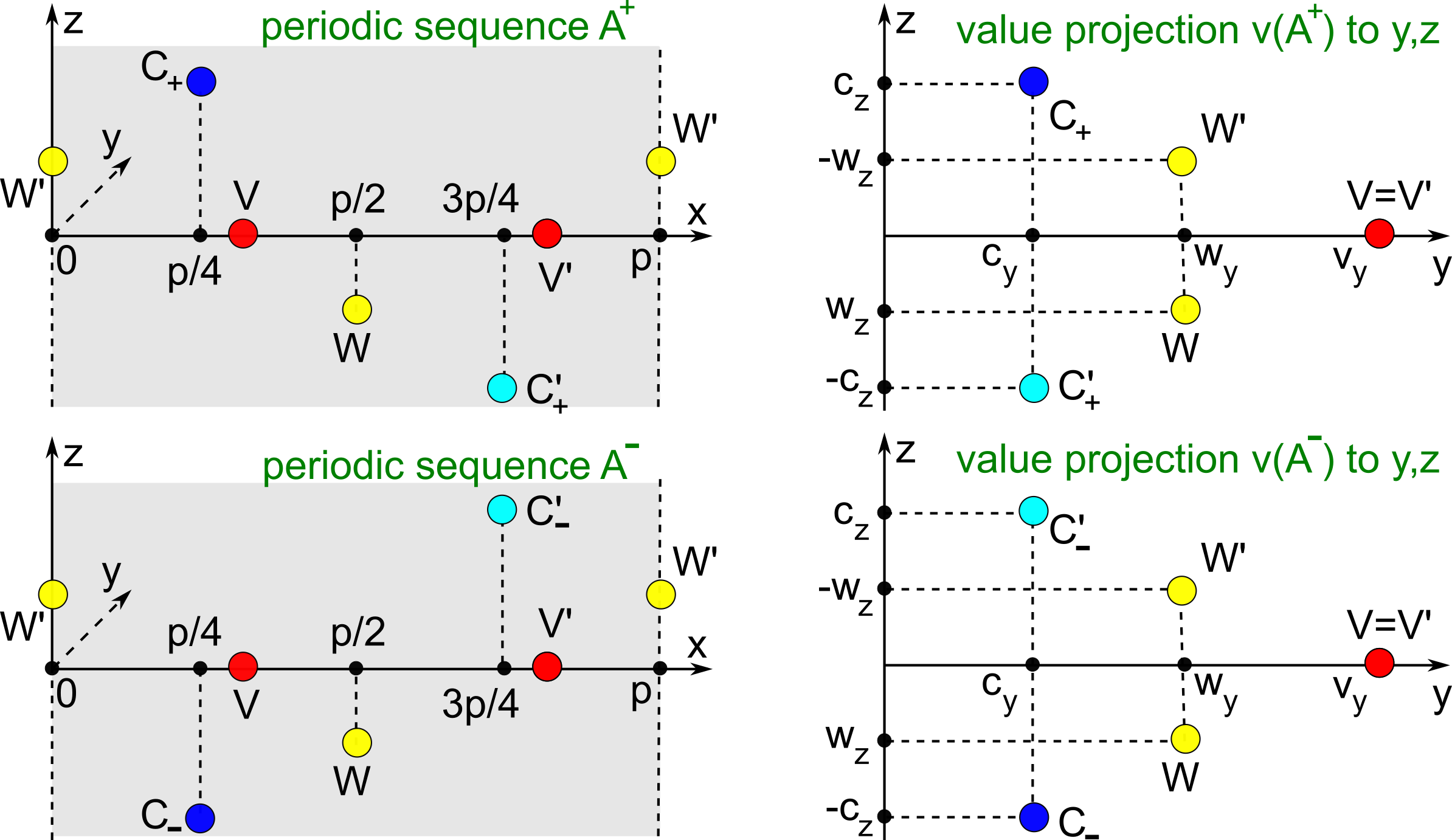}
\caption{These periodic sequences $A^{\pm}\subset\R\times\R^2$ from \cite[Fig.~2]{pozdnyakov2022incompleteness} have identical past invariants.}
\label{fig:6-point_periodic_sequences}
\end{figure}

Any point with a dash is obtained by $g(x,y,z)=(x+\frac{p}{2},y,-z)$.
The time projections are identical: $t(A^{\pm})=(0,\frac{p}{4},v_x,\frac{p}{2},\frac{3p}{4},\frac{p}{2}+v_x)$.
Assuming that $v_x\in(\frac{p}{4},\frac{p}{2})$ as in Fig.~\ref{fig:6-point_periodic_sequences}, the distance lists are $\DL^o(t(A^{\pm}))=(\frac{p}{4},v_x-\frac{p}{4},\frac{p}{2}-v_x,\frac{p}{4},v_x-\frac{p}{4},\frac{p}{2}-v_x)$.
The ordered value projections are $v(A^\pm)=\{(w_y,-w_z),(c_y,\pm c_z),(v_y,0),(w_y,w_z),(c_y,\mp c_z),(v_y,0)\}$.
The cyclic distance matrices of $A^+$ and $A^-$ on the left and right hand sides, respectively:
%By Definition~\ref{dfn:CDM} %cyclic distance matrices are 
%$\CDM(v(A^+))\neq\CDM(v(A^{-}))$ and produce a distance $\Elm^o>0$, which vanishes for $c_z=0$, when $A^{\pm}$ are actually isometric, so the latest pair has been distinguished.

\noindent
$\left(\begin{array}{llllll}
\mc{yellow}{d_{11}} & d_{12} & d_{21} & \mc{yellow}{d_{11}} & d_{12} & d_{21} \\
d_{21} & \mc{yellow}{d_{22}} & d_{12} & d_{21} & \mc{yellow}{d_{22}} & d_{12}\\
2|w_z| & 2|c_z|  & 0 & 2|w_z| & 2|c_z|  & 0
\end{array}\right)\neq
\left(\begin{array}{llllll}
\mc{yellow}{d_{22}} & d_{12} & d_{21} & \mc{yellow}{d_{22}} & d_{12} & d_{21} \\
d_{21} & \mc{yellow}{d_{11}} & d_{12} & d_{21} & \mc{yellow}{d_{11}} & d_{12}\\
2|w_z| & 2|c_z|  & 0 & 2|w_z| & 2|c_z|  & 0
\end{array}\right)$.
The differences are highlighted,
$d_{11}=\sqrt{(w_y-c_y)^2+(w_z\mc{yellow}{+c_z})^2}$,
$d_{12}=\sqrt{(c_y-v_y)^2+c_z^2}$, \\
$d_{22}=\sqrt{(w_y-c_y)^2+(w_z\mc{yellow}{-c_z})^2}$,
$d_{21}=\sqrt{(w_y-v_y)^2+w_z^2}$.
The matrix difference has the Minkowski norm $||\CDM(A^+)-\CDM(A^-)||_{\infty}=|d_{11}-d_{22}|>0$ unless $c_z=0$ or $w_z=0$.
If $c_z=0$, $A^{\pm}$ are identical.
If $w_z=0$, then $A^{\pm}$ are isometric by $g(x,y,z)=(x+\frac{p}{2},y,-z)$.
\medskip

If both $c_z,w_z\neq 0$, then $\Elm^o(A^+,A^-)$ is obtained by minimizing over 6 cyclic permutations $\si\in C^o(6)$. 
The trivial permutation and the shift by 3 positions give $|d_{11}-d_{12}|$.
Any other permutation gives $d_t=\max\{v_x-\frac{p}{4},\frac{p}{2}-v_x\}$ from comparing $\DL^o(t(A^+))$ with $\si(\DL^o(t(A^-)))$ and $d_v=\max\{|a-b|\}$ maximized for all pairs $a,b\in\{d_{11},d_{12},d_{21},d_{22}\}$.
In all cases, the elastic metric is positive: $\Elm^o(A^+,A^-)\geq|d_{11}-d_{22}|>0$.
Hence the time-value invariant $\TVI$ from Definitiion~\ref{dfn:TVI} distinguished this challenging pair $A^+\not\cong A^-$.
\bs
\end{exa}

Almost all properties in Theorem~\ref{thm:high-Elm} for high-dimensional elastic metrics will easily follow from similar properties in dimension $n=1$.
The only extra step is the continuity of the cyclic distance matrix $\CDM$ with respect to the maximum deviation of points below.

\begin{lem}[continuity of $\CDM$]
\label{lem:CDM}
Let ordered sets $S=\{p_1,\dots,p_m\}$ and $Q=\{q_1,\dots,q_m\}$  have a maximum deviation $\ep=\max\limits_{i=1,\dots,m}|p_i-q_i|$.
Then $||\CDM(S)-\CDM(Q)||_{\infty}\leq 2\ep$.
\bs
\end{lem}

\begin{thm}[solution to Problem~\ref{pro:metric} for high-dimensional  periodic sequences]
\label{thm:high-Elm}
For any periodic sequences $S,Q\subset\R\times\R^{n-1}$, let $m=\lcm(|S|,|Q|)$ be the lowest common multiple of their motif sizes.
The oriented elastic metric $\Elm^o(S,Q)$ can be computed in time $O(m^3)$, satisfies all metric axioms, and continuity $\Elm^o(S,Q)\leq 2d_B(S,Q)$ in the bottleneck distance when $2d_B(S,Q)$ is smaller than a minimum distance between all points of $t(S),t(Q)$. 
\bs
\end{thm}

%6========================
\section{Conclusions and a discussion of the impact of the new metrics}
%and further steps in Problem~\ref{pro:metric} for $n>1$}
\label{sec:discussion}

In conclusion, the key contribution is the introduction of easily computable and continuous elastic metrics in Definition~\ref{dfn:Elm} and \ref{dfn:high-Elm} on periodic sequences considered up to several equivalences.
Main Theorems~\ref{thm:complexity}, \ref{thm:continuity}, \ref{thm:TVI}, \ref{thm:high-Elm} resolved Problem~\ref{pro:metric} for 1-periodic sequences in any $\R^n$.
Even the case $n=1$ was open and required several new ideas such as comparing multiples of sequences with different periods using the Minkowski metric $||v||_{\infty}$. 
\medskip

Problem~\ref{pro:metric} is open for $k$-periodic sequences with $k\geq 2$ because conditions~(\ref{pro:metric}(d,e) on exact computability and continuity are the major obstacles.
For instance, Voronoi domains \cite{aurenhammer1991voronoi} are combinatorially unstable for finite point sets and even for 2-dimensional lattices.
\medskip

Example~\ref{exa:6-point_pair} was the major motivation for the new complete invariant $\TVI$ to distinguish all counter-examples developed in materials science \cite{pozdnyakov2022incompleteness}, where the past methods failed.
\medskip

The finite case of ordered sequences from section~\ref{sec:finite} has practical 
applications to proteins whose tertiary structures are polygonal backbones of ordered $\alpha$-carbon atoms embedded into $\R^3$.
One part of Proposition~\ref{prop:MCM} about the distance matrix $\DM$ might be folklore.
The distance $\MCM_q$ based on cyclic distance matrices was not applied to protein prediction. 
\medskip

Despite Google's AlphaFold \cite{senior2020improved,jumper2021highly} actually predicted inter-point distances between carbon atoms in a protein backbone, the final comparison between a predicted protein and  its `ground-truth' in the Protein Data Bank was done by the template modeling score.
%was a substantial improvement over past measures such as the Global Distance Test (GDT) and Root Mean Square Deviation (RMSD) between equivalent atoms in two proteins bu.
This TM-score \cite[formula~(1)]{zhang2004scoring} is $\TM=\max\left\{ \dfrac{1}{L_N}\sum\limits_{i=1}^{L_T} \dfrac{1}{1+(d_i/d_0)^2} \right\}\in[0,1]$ maximized over all spatial superpositions of two structures, where $\frac{d_i}{d_0}$ is a normalized distance between aligned residues or $\al$-carbon atoms, $L_T$ is the length of the template structure, $L_N$ is the length of a native structure.
Even if we forget about approximations caused by sampling superpositions, $1-\TM$ cannot be a metric already in simple cases for $L_T=L_N=1$ and $d_i/d_0$ taking pairwise distances $\frac{1}{2},\frac{1}{3},\frac{1}{4}$ between 3 atoms, which satisfy the triangle inequality, but $1-\TM$ takes the values $\frac{1}{5},\frac{1}{10},\frac{1}{17}$, which fail the triangle inequality. 
Hence future protein prediction can use proper metrics such as the easily computable metrics $\MCM_q$ in Proposition~\ref{prop:MCM}.
\medskip

A Python implementation of all new algorithms can be made public in October 2022.
\medskip

We thank all reviewers in advance for their valuable time and helpful suggestions.

%\end{document}

\appendix

%A========================
\section{Appendix: detailed proofs of all results}
\label{sec:isoset}

\begin{proof}[Proof of Lemma~\ref{lem:multiples}]
To prove that $\Elm^o(kS,Q)=\Elm^o(S,Q)$, let $m=\lcm(|S|,|Q|)$ and $m'=\lcm(k|S|,|Q|)$ be the lowest common multiples.
The proof for $\Elm$ is very similar.
\medskip

By Definition~\ref{dfn:Elm}, the elastic metric $\Elm^o(S,Q)$ is computed by comparing the sequences $\frac{m}{|S|}S$ and $\frac{m}{|Q|}Q$.
Since $\hat m=\frac{m'}{m}$ is integer, the metric $\Elm^o(kS,Q)$ uses the pair $\frac{m'}{|S|}S=\hat m\frac{m}{|S|}S$ and $\frac{m'}{|Q|}Q=\hat m\frac{m}{|Q|}Q$.
The pairs for $\Elm^o(S,Q)$ and $\Elm^o(kS,Q)$ differ by the factor of $\hat m$.
\medskip

The smallest distance lists $\SDL^o(\hat m\frac{m}{|S|}S)$ and $\SDL^o(\hat m\frac{m}{|Q|}Q)$ are obtained by concatenating $\hat m$ copies of $\SDL^o(\frac{m}{|S|}S)$ and $\SDL^o(\frac{m}{|Q|}Q)$, respectively.
For any cyclic permutation $\si\in C(\hat m m)$, the difference vector $\SDL^o(\hat m\frac{m}{|S|}S)-\si(\SDL^o(\hat m\frac{m}{|Q|}Q))$ is also a concatenation of $\hat m$ copies of $\SDL^o(\frac{m}{|S|}S)-\tau(\SDL^o(\frac{m}{|Q|}Q))$, where the cyclic permutation $\tau\in C(m)$ was obtained by restricting $\si\in C(\hat m m)$ to the first block of $m$ components, which is repeated $\hat m$ times in both long vectors. 
Since the Minkowski norm $||v||_{\infty}$ remains unchanged under concatenation of several copies of $v$, each resulting norm is the same, so $\Elm^o(kS,Q)=\Elm^o(S,Q)$.
\end{proof}

\begin{proof}[Proof of Lemma~\ref{lem:Elm}]
To check the axioms for $\Elm^o$, we use that 
the Minkowski norm $||v||_{\infty}$ defines a metric satisfying all axioms in Problem~\ref{pro:metric}(a,b,c).
The proof for $\Elm$ is similar.
\medskip
 
The first axiom requires that $\Elm^o(S,Q)=0$ if and only if $S,Q$ are related by translation.
The part \emph{if} follows from the invariance of $\SDL^o(S)$ in Lemma~\ref{prop:SDL}: if $S,Q$ are related by translation then $\SDL(S)=\SDL(Q)$.
The converse part \emph{only if} assumes that $\Elm^o(S,Q)=0$.
For $m=\lcm(|S|,|Q|)$, there is a cyclic permutation $\si\in C^o(m)$ such that $||\SDL^o(\frac{m}{|S|}S)-\si(\SDL^o(\frac{m}{|Q|}Q))||_{\infty}=0$.
Then the oriented smallest distance lists $\SDL^o(\frac{m}{|S|}S),\SDL^o(\frac{m}{|Q|}Q)$ coincide up to cyclic permutation.
The completeness of $\SDL^o$ in Lemma~\ref{prop:SDL} implies that the multiples $\frac{m}{|S|}S,\frac{m}{|Q|}Q$ are related by translation, then so are the original sequences $S,Q$.
\medskip
 
To check the symmetry axiom $\Elm^o(S,Q)=\Elm^o(Q,S)$, let $m=\lcm(|S|,|Q|)$.
Any cyclic permutation $\si\in C^o(m)$ has the inverse $\si^{-1}$ in the same group.
The vectors 
$\SDL^o(\frac{m}{|S|}S)-\si(\SDL^o(\frac{m}{|Q|}Q))$ and 
$\SDL^o(\frac{m}{|Q|}Q)-\si^{-1}(\SDL^o(\frac{m}{|S|}S))
=-(\si^{-1}(\SDL^o(\frac{m}{|S|}S))-\SDL^o(\frac{m}{|Q|}Q))$ 
differ by the cyclic permutation $\si$ and sign, hence have the same Minkowski norm.
Then the norm minimized over all cyclic permutations is the same, so
$\Elm^o(S,Q)=\Elm^o(Q,S)$.
\medskip

To check the triangle inequality for three periodic sequences $S,Q,T$,
let $m=\lcm(|S|,|Q|,|T|)$ be the lowest common multiple of their motif sizes.
By Lemma~\ref{lem:multiples}, the function $\Elm$ can be computed by using any multiple of a periodic sequence.
Hence we can use the periodic sequences $\frac{m}{|S|}S, \frac{m}{|Q|}Q,\frac{m}{|T|}T$, which all have the same number $m$ of points in their motifs.
Only for simplicity, we denote these multiple sequences by the same letters $S,Q,T$ below.
\medskip

In Definition~\ref{dfn:Elm}, let $\si\in C^o(m)$ and $\tau\in C^o(m)$ be cyclic permutations that minimize $\Elm^o(S,Q)=||\SDL^o(S)-\si(\SDL^o(Q))||_{\infty}$ and $\Elm^o(Q,T)=||\SDL^o(Q)-\tau(\SDL^o(T))||_{\infty}$, respectively.
We use the composition $\si\circ\tau$ and the triangle inequality for $||v||_{\infty}$:
$\Elm^o(S,T)\leq$ 
$||\SDL^o(S)-\si\circ\tau(\SDL^o(T))||_{\infty}=
||\SDL^o(S)-\si(\SDL^o(Q))+\si(\SDL^o(Q)-\tau(\SDL^o(T)))||_{\infty}
\leq$
$||\SDL^o(S)-\si(\SDL^o(Q))||_{\infty}+||\si(\SDL^o(Q)-\tau(\SDL^o(T)))||_{\infty}=\Elm^o(S,Q)+\Elm^o(Q,T).$
\end{proof}

\begin{proof}[Proof of Theorem~\ref{thm:complexity}]
For ordered motif points $p_1,\dots,p_{|S|}$ within a period interval $[0,l)$ of $S$, in time $O(|S|)$, compute the list $\DL(S)$ of distances $d_i=p_{i+1}-p_i$ for $i=1,\dots,|S|$, where $p_{|S|+1}=p_1+l$.
The same distance list $\DL(Q)$  similarly needs $O(|Q|)$ time.
\medskip

By Definition~\ref{dfn:Elm} of elastic metrics, find the lowest common multiple $m=\lcm(|S|,|Q|)$.
The sequences $\frac{m}{|S|}S$ and $\frac{m}{|Q|}Q$ have distance lists $\DL(\frac{m}{|S|}S),\DL(\frac{m}{|Q|}Q)$ obtained by concatenating $\frac{m}{|S|},\frac{m}{|Q|}$ copies of $\DL(S),\DL(Q)$, respectively, which uses only $O(m)$ time.
\medskip

To compute $\Elm^o(S,Q)$ and $\Elm(S,Q)$ by Definition~\ref{dfn:Elm}, we can use the above lists $\DL(\frac{m}{|S|}S),\DL(\frac{m}{|Q|}Q)$ without finding the smallest (up to cyclic permutations) distance lists because the elastic metrics are minimized over cyclic permutations anyway.
\medskip

For each of $O(m)$ cyclic permutations $\si$ for both metrics $\Elm^o$ and $\Elm$, we find the minimum norm $||\DL(\frac{m}{|S|}S)-\si(\DL(\frac{m}{|Q|}Q))||_{\infty}$ in time $O(m)$, so the overall time is $O(m^2)$.
\end{proof}

\begin{proof}[Proof of Theorem~\ref{thm:continuity}]
By Lemma~\ref{lem:multiples} one can replace $S,Q$ by their multiples without changing the elastic metrics.
If we need to consider multiples of $S,Q$ to compute $\Elm^o(S,Q)$ by Definition~\ref{dfn:Elm}, we continue using the same symbols $S,Q$ for simplicity because the bottleneck distance $d_B$ also satisfies Lemma~\ref{lem:multiples}.
So we can assume that both $S=\{p_1,\dots,p_m\}+l\Z$ and $Q=\{q_1,\dots,q_m\}+l'\Z$ have $m$ motif points, though their periods $l,l'$ can differ. 
\medskip

In Definition~\ref{dfn:bottleneck} of the bottleneck distance $d_B$, let $\ep=d_B(S,Q)$ and $h:S\to Q$ be an optimal bijection such that $|p-h(p)|\leq\ep$ for any point $p\in S$.
Consider two cases.
\medskip

\emph{Simple case}:
first assume that both $S,Q$ have a minimum inter-point distance more than $2d_B(S,Q)$.
When we move any point $p_i\in S$ to its bijective image $h(p_i)\in Q$ by a distance (at most $\ep$) smaller than a half of all inter-point distances, any successive points $p_i,p_{i+1}\in S$ remain successive without colliding with other points of $S$ for $i=1,\dots,m$, where all indices are considered modulo $m$, so $p_{m+1}=p_1+l$.
By cyclically renumbering the points of $Q$, one can assume that $h(p_i)=q_i$ for $i=1,\dots,m$.
Since the points in each pair $p_i,q_i$ are $\ep$-close, every distance $d_i(S)=p_{i+1}-p_{i}$ differs from $d_i(Q)=q_{i+1}-q_i$ by at most $2\ep$.
\medskip

\emph{General case}: the bijection $h:S\to Q$ may not respect the cyclic order of points.
For any successive  points of $S$, the open interval $I=(p_i,p_{i+1})$ is called \emph{long} if its length $|p_{i+1}-p_i|>2\ep$, otherwise \emph{short}.
Since every point $p_i\in S$ moves under $h$ to its image $h(p_i)\in Q$ by less than $\ep$, any long interval $I$ can become shorter or longer by at most $2\ep$.
Let $p_j\leq p_i$ is the point with a maximum difference $h(p_j)-h(p_i)\in[0,\ep)$ so that $p_j$ `jumps over' $p_i$ under $h$ from the left to the most distant position on the right of $p_i$ (the case $i=j$ is possible).  
Similarly, find the point $p_k\geq p_{i+1}$ that `jumps over' $p_{i+1}$ under $h$ (by at most $\ep$) from the right to the most distant position on the left of $p_{i+1}$ ($i+1=k$ is possible). 
\medskip

Hence any long interval $I=(p_i,p_{i+1})$ of $S$ has a well-defined interval $h^*(I)=(h(p_j),h(p_k))$ between the successive points of $Q$ with a length more than $|p_{i+1}-p_i|-2\ep>0$.
\medskip

The endpoints of all short intervals between any two successive intervals $I,I'$ of $S$ under $h$ remain between the images $h^*(I),h^*(I')$ of $Q$ because these endpoints cannot `jump over' the initial long intervals $I,I'$ of lengths more than $2\ep$.
Then $h^*$ extends to a bijection from all ordered short intervals between $I,I'$ to all ordered intervals between $h^*(I),h^*(I')$.
\medskip

So $h^*$ becomes a bijection between all intervals (between successive points) in $S$ and $Q$, respecting the cyclic order of these intervals.
It remains to show that any interval $I$ of $S$ changes its length under $h^*$ by at most $2\ep$. 
The above proof of this fact for any long interval $I$ of $S$ similarly works for any long interval $h^*(I)$ of $Q$.
If both interval $I$ and its image $h^*(I)$ are short (not longer than $2\ep$), their lengths clearly differ at most $2\ep$.  
\medskip

Both cases above concluded that (up to a cyclic permutation) all corresponding coordinates in the distance lists $(d_1(S),\dots,d_m(S))$ and $(d_1(Q),\dots,d_m(Q))$ differ by at most $2\ep$.
By minimizing for all cyclic permutations in Definition~\ref{dfn:Elm}, we  can get only a smaller Minkowski norm $||v||_{\infty}$ for the difference $v$ of the lists above.
Hence $\Elm^o(S,Q)\leq 2\ep$.  
\medskip

The inequality $\Elm(S,Q)\leq\Elm^o(S,Q)$ follows by Definition~\ref{dfn:Elm} because $\Elm(S,Q)$ is minimized over the larger group of permutations in comparison with $\Elm^o(S,Q)$.
\end{proof}

\begin{proof}[Proof of Proposition~\ref{prop:MCM}]
Since both matrices $\DM(T)$ and $\CDM(T)$ consist of the same non-zero distances up to a cyclic shift in every column, the Minkowski metric gives the same function: $||\DM(S)-\DM(T)||_q=||\CDM(S)-\CDM(T)||_q$ for any $q\in[1,+\infty]$.
\medskip

The symmetry and triangle inequality for $\MCM_q$ follow from the same axioms of the Minkowski metric.
The first axiom means that $\CDM(T)$ is a complete isometry invariant.
\medskip

Any isometry preserving the order of points makes the matrices $\CDM(S)=\CDM(T)$ equal element-wise.
Conversely, given a matrix $\CDM(T)$ from Definition~\ref{dfn:CDM}, one can uniquely reconstruct $T=\{p_1,\dots,p_m\}$ up to isometry in $\R^{n}$ as follows.
Fix $p_1$ at the origin of $\R^{n}$.
\medskip

The distance $\CDM_{11}(T)=|p_1-p_2|$ allows us to fix $p_2$ in the (positive) $x_1$-coordinate axis of $\R^{n}$, while we can rotate the whole set $T$ around the $x_1$-axis.
The distances $\CDM_{12}(T)=|p_2-p_3|$ and $\CDM_{21}(T)=|p_1-p_3|$ determine $p_3$ in the $(x_1,x_2)$-plane of $\R^{n}$ and we can isometrically map $T$ by $f\in\Or(\R^{n})$ preserving this plane. 
If $p_1,p_2,p_3$ are in one line, the position of $p_3$ in the $x_1$-axis is still unique and we can rotate $T$ around the $x_1$-axis. 
\medskip

We continue using further distances to similarly fix all the points $p_1,\dots,p_m\in\R^n$.
If the full sequence $T$ affinely generates a smaller subspace $\R^k\subset\R^{n}$ for $k<n$, this construction determines $T\subset\R^k$ up to isometry of $\R^k$, hence up to isometry of $\R^{n}$.
\end{proof}

\begin{proof}[Proof of Theorem~\ref{thm:TVI}]
By Definition~\ref{dfn:high-dim_sequence} any equivalence $f\times g$ acts by translations $f$ on the time projection, hence preserves the distance list $\DL^o(t(S))$ up to cyclic permutation.
For a fixed time projection of $m$ ordered points, the isometry $g$ preserves the cyclic distance matrix $\CDM(v(S))$.
Given $\TVI(S)=(\DL^o(t(S)),\CDM(v(S)))$, we reconstruct the sequence $t(S)$ up to translation in $\R$ by Proposition~\ref{prop:SDL} and the ordered set $v(S)$ up to isometry in $\R^{n-1}$ by Proposition~\ref{prop:MCM}.
Then $S$ is determined up to equivalence by $t(S)$ and $v(S)$.
\end{proof}

\begin{proof}[Proof of Lemma~\ref{lem:CDM}]
The triangle inequality for the Euclidean distance implies that the corresponding elements of the cyclic distance matrices from Definition~\ref{dfn:CDM} differ by at most $2\ep$ as follows: $|d_{ij}(S)-d_{ij}(Q)|=||p_j-p_{i+j}|-|q_j-q_{i+j}||\leq |p_i-q_j|+|p_{i+j}-q_{i+j}|\leq 2\ep$.
\end{proof}

\begin{proof}[Proof of Theorem~\ref{thm:high-Elm}]
The proof is similar to the case $n=1$.
For the time complexity, the extra factor $m\geq n$ emerges because the cyclic distance matrices $\CDM(S),\CDM(Q)$ have at most $m\times m$ distances.
Computing the Minkowski norm of the matrix difference needs $O(m^2)$ operations for each of $m$ cyclic permutations to find the metric $\Elm^o$ by Definition~\ref{dfn:high-Elm}. 
\medskip

To check the metric axioms, we start from the analog of Lemma~\ref{lem:multiples} implying that the elastic metric $\Elm^o$ is independent of a non-minimum period also for high-dimensional sequences.
The first axiom follows from Propositions~\ref{prop:SDL} and~\ref{prop:MCM} because $\Elm^o(S,Q)=0$ implies that there is a cyclic permutation $\si\in C^o(m)$ matching $\DL^o(S)$ with $\DL^o(Q)$ and $\CDM(S)$ with $\CDM(Q)$, hence making the sequences $S,Q$ equivalent by Definition~\ref{dfn:high-dim_sequence}.
\medskip

The symmetry axiom easily follows similarly to the proof of Lemma~\ref{lem:Elm} because any cyclic permutation $\si\in C^o(m)$ has its inverse $\si^{-1}$ in the same group.
To prove the triangle inequality, we find multiples of $S,Q,T$ that have the same number $m$ of motif points.
We continue denoting these sequences by $S,Q,T$ for simplicity.
Let $\si,\tau\in C^o(m)$ be cyclic permutations that minimize $\Elm^o(S,Q)$ and $\Elm^o(Q,T)$, respectively, in Definition~\ref{dfn:high-Elm}.
Then $d_t=||\DL^o(S)-\si\circ\tau(\DL^o(T))||_{\infty}\leq 
||\DL^o(S)-\si(\DL^o(Q))||_{\infty}+||\si(\DL^o(Q)-\tau(\DL^o(T)))||_{\infty}\leq\Elm^o(S,Q)+\Elm^o(Q,T)$.
Since the above inequalities hold for $d_v$ after replacing $\DL^o$ by $\CDM$, we get $\Elm^o(S,T)=\max\{d_t,d_v\}\leq\Elm^o(S,Q)+\Elm^o(Q,T)$.
The final inequality $\Elm^o(S,Q)\leq 2d_B(S,Q)$ directly follows from Theorem~\ref{thm:continuity} and Lemma~\ref{lem:CDM}.
\end{proof}

\bibliography{metric-sequences}

\end{document}